\documentstyle[prc,aps,preprint]{revtex}
\begin{document} \draft
\date{\today}
\title{Coupling constants g$_{a_0\omega\gamma}$  and g$_{a_0\rho\gamma}$
as derived from QCD sum rules}

\author{A. Gokalp~\thanks{agokalp@metu.edu.tr} and
        O. Yilmaz~\thanks{oyilmaz@metu.edu.tr}}
\address{ {\it Physics Department, Middle East Technical University,
06531 Ankara, Turkey}}
\maketitle

\begin{abstract}
We consider the two point correlation function of scalar current
in QCD sum rules approach to estimate the overlap amplitude of
$a_0$ meson. We then employ QCD sum rules to calculate the
coupling constants g$_{\omega a_0\gamma}$ and g$_{\rho a_0\gamma}$
by studying the three point ${a_0\omega\gamma}$- and
${a_0\rho\gamma}$-correlation functions.
\end{abstract}

\thispagestyle{empty} ~~~~\\ 
\newpage
\setcounter{page}{1}
\section{Introduction}

The low-mass scalar mesons have fundamental importance in
understanding the theory and phenomenology of low energy QCD. From
the experimental point of view, isoscalar $f_0(980)$ and isovector
$a_0(980)$ are well established, but the nature and the quark
substructure of these scalar mesons, whether they are conventional
$q\overline{q}$ states \cite{R1}, $K\overline{K}$ molecules
\cite{R2}, or multiquark exotic $q^2\overline{q}^2$ states
\cite{R3}, have been a subject of controversy. On the other hand,
since they are relevant hadronic degrees of freedom, besides the
questions of their nature, the roles of scalar mesons in the
hadronic processes must be studied.

The radiative decay processes of the type $V^{0}\rightarrow
P^{0}P^{0}\gamma$ where V and P belong to the lowest multiplets of
vector (V) and pseudoscalar (P) mesons have become a subject of
renewed interest because they offer the possibility of
investigating the new physics features governing meson physics in
the low energy region. Although these rare decays have small
branching ratios due to absence of bremsstrahlung radiation, their
study offers the possibility of testing the theoretical ideas
about the interesting mechanisms of these decays, as well as
shedding light on the structure of intermediate states involved in
these decays. Particularly interesting are the exchange mechanisms
of scalar resonances contributing to these decays. The radiative
decays $\rho^{0}\rightarrow\pi^{0}\eta\gamma$ and
$\omega\rightarrow\pi^{0}\eta\gamma$ were studied using a low
energy effective Lagrangian approach with gauged Wess-Zumino terms
\cite{R4}, and later by using standard Lagrangians obeying
SU(3)-symmetry \cite{R5}. In both of these calculations scalar
meson intermediate state contributions were neglected and the
contributions of intermediate vector mesons were taken into
account. However, it is of interest to study the contribution of
$a_0$-intermediate state to these decays as well, and for that a
knowledge of $a_0\omega\gamma$- and $a_0\rho^0\gamma$-vertexes are
needed.

In this work, we estimate the coupling constant
g$_{a_0\rho\gamma}$  and g$_{a_0\omega\gamma}$ by employing QCD
sum rules which provide an efficient method to study hadronic
properties and which have been employed to study hadronic
observables such as decay constants and form factors in terms of
nonperturbative contributions proportional to the quark and gluon
condensates \cite{R6,R7,R8}.

\section{Analysis and Results}

The QCD sum rules approach  \cite{R6,R7,R8} is a model independent
method to study the properties of hadrons through correlation
functions of appropriately chosen currents. We choose the
interpolating currents for $\omega$ and $\rho$ mesons as
$j_{\mu}^\omega=\frac{1}{2}(\overline{u}\gamma
u+\overline{d}\gamma d)$ and
$j_{\mu}^\rho=\frac{1}{2}(\overline{u}\gamma u-\overline{d}\gamma
d)$ respectively; and for $a_0$ meson as
$j_{a_0}=\frac{1}{2}(\overline{u}u-\overline{d}d)$ \cite{R6,R7},
and we work in the SU(2) flavour limit m$_{u}=m_{d}=m_{q}$. In the
sum rule, the overlap amplitude of $a_0$ meson
$\lambda_{a_0}=<0|j_{a_0}|a_0>$ is needed. In a previous work
\cite{R9} we studied the scalar-isoscalar $\sigma$ meson by
considering the two-point scalar current correlation function.
Since perturbative and QCD-vacuum condensate contributions to
scalar current correlation functions cannot distinguish between
isoscalar and isovector channels, we follow here the same method
and we study the scalar-isovector $a_0$ meson by considering the
two-point current correlation function
\begin{equation}\label{e1}
  \Pi(p^{2})=i\int d^{4}x
  e^{ip.x}<0|T\{j_{a_0}(x)j^\dag_{a_0}(0)\}|0>~~.
\end{equation}
The two-loop expression for the scalar current correlation
function $\Pi(p^2)$ in perturbative QCD was calculated \cite{R10},
and for light quark systems in the limit $m_q=0$ it is given by
the expression
\begin{equation}\label{e2}
  \Pi_{pert}(p^{2})=\frac{3}{16\pi^2}(-p^2)\ln
  (\frac{-p^2}{\mu^2})\left\{ 1+\frac{\alpha_{s}}{\pi}\left [\frac{17}{3}-\ln
  (\frac{-p^2}{\mu^2})\right ]\right\}~~.
\end{equation}
QCD-vacuum condensate  contributions to the scalar current
correlation function $ \Pi(p^{2})$ were obtained by the operator
product method  \cite{R11} in the same limit $m_q$=0 as
\begin{equation}\label{e3}
  \Pi(p^{2}=-Q^2)_{cond}=\frac{3}{2Q^2}<m_q\overline{q}q>
  +\frac{1}{16\pi Q^2}<\alpha_s G^2>-
  \frac{88\pi}{27 Q^4}<\alpha_s(\overline{q}q)^2>~~.
\end{equation}
Let us note that the term $<m_q\overline{q}q>$ is independent of
quark mass since it is given as $-f_{\pi}^{2}m_{\pi}^{2}/4$
through Gell-Mann-Oakes-Renner relation \cite{R6}.

The correlation function $ \Pi(p^{2})$ satisfies the standard
subtracted dispersion relation \cite{R6}
\begin{equation}\label{e4}
  \Pi_{pert}(p^{2})=p^2\int_0^\infty\frac{ds}{s(s-p^2)}\rho(s)+\Pi(0)
\end{equation}
where the spectral density function is given as
$\rho(s)=(1/\pi)Im\Pi(s)$. The spectral density contains a single
sharp pole $\pi\lambda_{a_0}\delta(s-m_{a_0}^2)$ corresponding to
the coupling of $a_0$ meson to the scalar current. The continuum
contribution of the higher states to the spectral density is
estimated as $\rho=\rho_h(s)\theta(s-s_0)$ where s$_0$ denote the
continuum threshold and $\rho_h$ is given by the expression
$\rho_h(s)=(1/\pi)Im \Pi_{OPE}(s)$ with $ \Pi_{OPE}(s)$ is
obtained from Eq. (2) and Eq. (3) as
$\Pi_{OPE}(s)=\Pi_{pert}(s)+\Pi_{cond}(s)$. After performing the
Borel transformation we obtain the QCD sum rule for the overlap
amplitude  $\lambda_{a_0}$ as
\begin{eqnarray}\label{e5}
  \lambda_{a_0}e^{-\frac{m_{a_0}^2}{M^2}}=\frac{3}{16\pi^2}M^2\left\{
  \left [1-(1+\frac{s_0}{M^2})e^{\frac{-s_0}{M^2}}\right ]
  \left(1+\frac{\alpha_{s}(M)}{\pi}\frac{17}{3}\right)-
  2\frac{\alpha_{s}(M)}{\pi}\int_0^{s_0/M^2} w\ln we^{-w}dw\right\}
 \nonumber \\
  + \frac{3}{2M^2}<m_q\overline{q}q>+\frac{1}{16\pi M^2}<\alpha_s G^2>-
  \frac{88\pi}{27 M^4}<\alpha_s(\overline{q}q)^2>~~.
\end{eqnarray}
In the numerical evaluation of Eq. (5) we use
$<m_q\overline{q}q>=(-0.82\pm 0.10)\times 10^{-4}~~GeV^4$,
$<\alpha_s G^2>=(0.038\pm 0.011)~~GeV^4$,
$<\alpha_s(\overline{q}q)^2>=-0.18\times 10^{-3}~~GeV^6$
\cite{R8,R12}. The threshold is choosen below a possible
$a_0(1450)$ pole and it is varied between  s$_0$=1.6-1.7 GeV$^2$.
Since the Borel parameter has no physical meaning, we look for a
range of its values where the sum rule is almost independent of
M$^2$, we choose the interval of values of Borel parameter M$^2$
as 1.2-2.0 GeV$^2$. The overlap amplitude $\lambda_{a_0}$ as a
function of M$^2$ in this interval for different values of s$_{0}$
is shown in Fig. 1 from which by choosing the middle value
M$^2$=1.6 GeV$^2$ in it is interval of variation, we obtain the
overlap amplitude as $\lambda_{a_0}=0.21\pm 0.05~~GeV^2$ where we
include the uncertainty due to the variation of the continuum
threshold and the Borel parameter $M^2$ as well as the uncertainty
due to errors attached to the estimated values of condensates as
quoted above.

In order to derive the QCD sum rule for the coupling constants
g$_{a_0\omega\gamma}$ and g$_{a_0\rho\gamma}$, we consider the
three point correlation function
\begin{equation}\label{e6}
  T_{\mu\nu}(p,p^\prime)=\int d^{4}x d^4y e^{ip^\prime.y}e^{-ip.x}
  <0|T\{j_\mu^\gamma(0)j_{\nu}^V(x)j_{a_0}(y)\}|0>~~,
\end{equation}
where
$j_{\mu}^{\gamma}=(e_u\overline{u}\gamma_{\mu}u+e_d\overline{d}\gamma_{\mu}d)$
is the electromagnetic current with $e_u$ and $e_d$ being the
quark charges, and $j_{\nu}^{V}$ is the interpolating current for
$\omega$ or $\rho^0$ meson.

In order to obtain the phenomenological part of the sum rule, we
consider the double dispersion relation for the vertex function
$T_{\mu\nu}$
\begin{equation}\label{e7}
  T_{\mu\nu}(p,p^\prime)=\frac{1}{\pi^2}\int ds_1\int ds_2
  \frac{\rho_{\mu\nu}(s_1,s_2)}{(p^2-s_1)({p^{\prime}}^2-s_2)}~~,
 \end{equation}
where the possible subtraction terms are neglected since they will
not make any contribution after double Borel transform. For low
values of $s_1$ and $s_2$, the spectral function
$\rho_{\mu\nu}(s_1,s_2)$ contains a term proportional to double
$\delta$-function $\delta(s_1-m_V^2)\delta(s_2-m_{a_0}^2)$,
corresponding to the transition $a_0\rightarrow V\gamma$ where V
denotes $\omega$ or $\rho^0$ meson. We therefore saturate the
dispersion relation satisfied by the vertex function $T_{\mu\nu}$
by these lowest lying meson states in the vector and the scalar
channels, and this way we obtain for the physical part
\begin{equation}\label{e8}
  T_{\mu\nu}(p,p^\prime)=\frac{<0|j_{\nu}^V|V>
  <V(p)|j_\mu^\gamma|a_0(p^\prime)><a_0|j_{a_0}|0>}
  {(p^2-m^2_V)({p^\prime}^2-m^2_{a_0})}+...
\end{equation}
where the contributions from the higher states and the continuum
is denoted by dots. In this expression the overlap amplitude
$\lambda_{a_0}=<a_0|j_{a_0}|0>$ of $a_0$-meson has been determined
in previous sections. The overlap amplitude $\lambda_V$ of vector
meson is defined as $<0|j_{\nu}^V|V>=\lambda_V u_\nu$ where
$u_\nu$ is the polarization vector of the vector meson $\omega$ or
$\rho^0$. The matrix element of the electromagnetic current is
given as
\begin{equation}\label{e9}
<V(p)|j_\mu^\gamma|a_0(p^\prime)>=
-i\frac{e}{m_V}g_{a_0V\gamma}K(q^2)(p\cdot q ~u_\mu -u\cdot q~
p_\mu)~~,
\end{equation}
where $q=p-p^\prime$ and $K(q^2)$ is a form factor with K(0)=1.
This expression defines the coupling constant through the
effective Lagrangian
\begin{equation}\label{e10}
{\cal L}=\frac{e}{m_V}g_{a_0V\gamma}\partial^\alpha
V^\beta(\partial_\alpha A_\beta-\partial_\beta A_\alpha )a_0
\end{equation}
describing the $a_0V\gamma$-vertex.

The theoretical part of the sum rule is obtained by calculating
the perturbative contribution and the power corrections from
operators of different dimensions to the three point correlation
function $T_{\mu\nu}$. For the perturbative contribution we
consider the lowest order bare-loop diagrams shown in Fig. 2(a).
Furthermore, we consider the power corrections from the operators
of different dimensions that are proportional to vacuum
condensates $<\overline{q}q>$, $<\overline{q}\sigma\cdot G q>$ and
$<(\overline{q}q)^2>$. We do not consider the gluon condensate
contribution proportional to $<G^2>$ since it is estimated to be
negligible for light quark systems. We perform the calculations of
the power corrections in the fixed point gauge \cite{R13}. We work
in the limit $m_q=0$, and in this limit perturbative bare-loop
diagram does not make any contribution. Moreover, in this limit
only operators of dimensions d=3 and d=5 make contributions that
are proportional to $<\overline{q}q>$ and
$<\overline{q}\sigma\cdot G q>$, respectively. The relevant
Feynman diagrams for power corrections are shown Fig. 2(b) and
(c). If we consider the gauge invariant structure ($p_\mu q_\nu
-p\cdot q g_{\mu\nu}$), we obtain the power corrections of
dimensions d=3 and d=5 as
\begin{equation}\label{e11}
  C_3=i\frac{3}{4}\frac{1}{p^2}\frac{1}{{p^\prime}^2}(e_u<\overline{u}u>+
e_d<\overline{d}d>)
\end{equation}
and
\begin{equation}\label{e12}
  C_5=\left(i\frac{9}{32}\frac{1}{p^4}\frac{1}{{p^\prime}^2}+
  i\frac{1}{32}\frac{1}{p^2}\frac{1}{{p^\prime}^4}\right)
  (e_u<g_s\overline{u}\sigma\cdot G u>+e_d<g_s\overline{d}\sigma\cdot G
  d>)~~.
\end{equation}

After performing double Borel transform with respect to the
variables $Q^2=-p^2$ and ${Q^\prime}^2=-{p^\prime}^2$, and by
considering the gauge-invariant structure ($p_\mu q_\nu -p\cdot q
g_{\mu\nu}$) for the phenomenological part as well, we obtain the
sum rule for the coupling constant g$_{a_0V\gamma}$
\begin{eqnarray}\label{e13}
g_{a_0V\gamma}=-e_q<\overline{u}u>
\frac{3m_V}{\lambda_{a_0}\lambda_V}
  e^{\frac{m_{a_0}^2}{M^2}}e^{\frac{m_V^2}{{M^\prime}^2}}
    \left (\frac{3}{4}-\frac{9}{32}m_0^2\frac{1}{M^2}
    -\frac{1}{32}m_0^2\frac{1}{{M^\prime}^2}\right )
\end{eqnarray}
where $e_q=(e_u+e_d)$ for $\rho^0$ meson and $e_q=(e_u-e_d)$ for
$\omega$ meson, and  we use the relations
$<\overline{q}\sigma\cdot G q>=m_0^2<\overline{q}q>$ and
$<\overline{u}u>=<\overline{d}d>$. For the numerical evaluation of
the sum rule we use the values $m_0^2=(0.8\pm 0.02)~~GeV^2$,
$<\overline{u}u>=(-0.014\pm 0.002)~~GeV^3$ \cite{R8,R14}, and
$m_\rho=0.770~~GeV$, $m_\omega=0.782~~GeV$. For the overlap
amplitude $\lambda_{a_0}$ we use the value $\lambda_{a_0}=(0.21\pm
0.05)~~GeV^2$ that we have estimated previously. We determine the
overlap amplitude $\lambda_V$ for $\omega$ and $\rho^0$ meson from
the measured leptonic decay widths $\Gamma(V\rightarrow e^+e^-)$
\cite{R15}, thus we use their experimental values
$\lambda_\rho=(0.117\pm 0.003)~~GeV^2$  and
$\lambda_\omega=(0.108\pm 0.002)~~GeV^2$. In order to analyze the
dependence of g$_{a_0V\gamma}$ on Borel parameters $M^2$ and
${M^\prime}^2$, we study the independent variations of $M^2$ and
${M^\prime}^2$ in the interval $0.6~~GeV^2\leq
M^2,{M^\prime}^2\leq 1.4~~GeV^2$ as these limits determine the
allowed interval for the vector channel \cite{R16}. The variation
of the coupling constant g$_{\omega a_0\gamma}$ as a function of
Borel parameter $M^2$ for different values of ${M^\prime}^2$ is
shown in Fig. 3, examination of which indicates that it is quite
stable with these reasonable variations of $M^2$ and
${M^\prime}^2$. We choose the middle value $M^2=1~~GeV^2$ for the
Borel parameter in its interval of variation and we obtain the
coupling constant g$_{a_0\omega\gamma}$ as g$_{a_0\omega
\gamma}=(0.75\pm 0.20)$. We indicate the error arising from the
numerical analysis of the sum rule as well as from the
uncertainties in the estimated values of the vacuum condensates.
In Fig. 4 we present the variation of the coupling constant
g$_{a_0\rho\gamma}$ as a function of the Borel parameter $M^2$ for
different values of ${M^\prime}^2$. Following a similar analysis
as in the case of g$_{a_0\omega\gamma}$, we obtain the coupling
constant g$_{a_0\rho\gamma}$ as g$_{a_0\rho\gamma}=(2.00\pm
0.50)$. The values for the coupling constants
g$_{a_0\omega\gamma}$ and g$_{a_0\rho\gamma}$ that we obtain are
in agreement with the expected SU(3) ratio
g$_{a_0\rho\gamma}$:g$_{a_0\omega\gamma}$=3:1.

In our analysis, we use for the overlap amplitudes
$\lambda_\omega$ and $\lambda_\rho$ the values that we obtain from
the experimental electronic decay widths of $\omega$ and $\rho^0$
mesons. On the other hand, it may be argued that in a QCD sum
rules calculation it is more appropriate to use the values of the
overlap amplitudes that are also determined within the framework
of QCD sum rules method. Electromagnetic decays of vector mesons
using QCD sum rules were studied \cite{R17}, and in their analysis
the authors used the values of the overlap amplitudes
$\lambda_\omega=(0.16\pm 0.01)~~GeV^2$ and $\lambda_\rho=(0.17\pm
0.01)~~GeV^2$ that they also determined utilizing QCD sum rules.
If we use instead these values of the overlap amplitudes in our
calculation we then obtain the coupling constants
g$_{a_0\omega\gamma}$ and g$_{a_0\rho\gamma}$ as
g$_{a_0\omega\gamma}=0.45\pm 0.10$ and g$_{a_0\rho\gamma}=1.30\pm
0.30$ which are consistent with our above results.

In the investigations of the role of $a_0$ meson in hadronic
processes, the relevant coupling constants of $a_0$ meson are
needed. In this work, we employed QCD sum rules approach to
estimate the coupling constants g$_{a_0\omega\gamma}$ and
g$_{a_0\rho\gamma}$. We feel that the studies of the different
coupling constants of $a_0$ meson should be continued. In
particular QCD sum rules calculations should be improved by taking
into account the high order corrections to the perturbative part
of the three point correlation function and also to the two point
correlation function employed in the estimation of the overlap
amplitude of $a_0$ meson.


\newpage

{\bf Figure Captions:}

\begin{description}

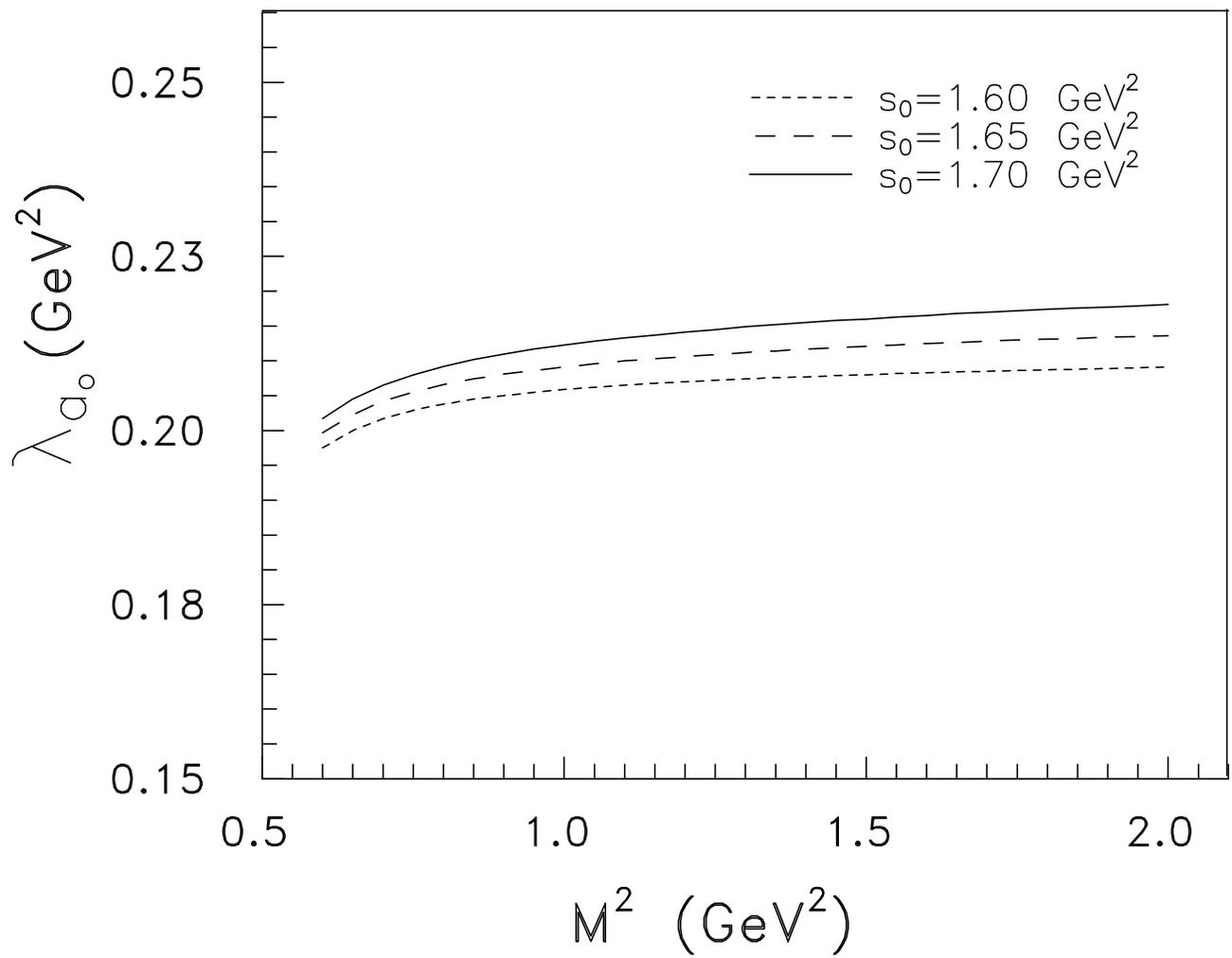
\item[{\bf Figure 1}:] The overlap amplitude $\lambda_{a_0}$ as a
function of Borel parameter $M^2$.

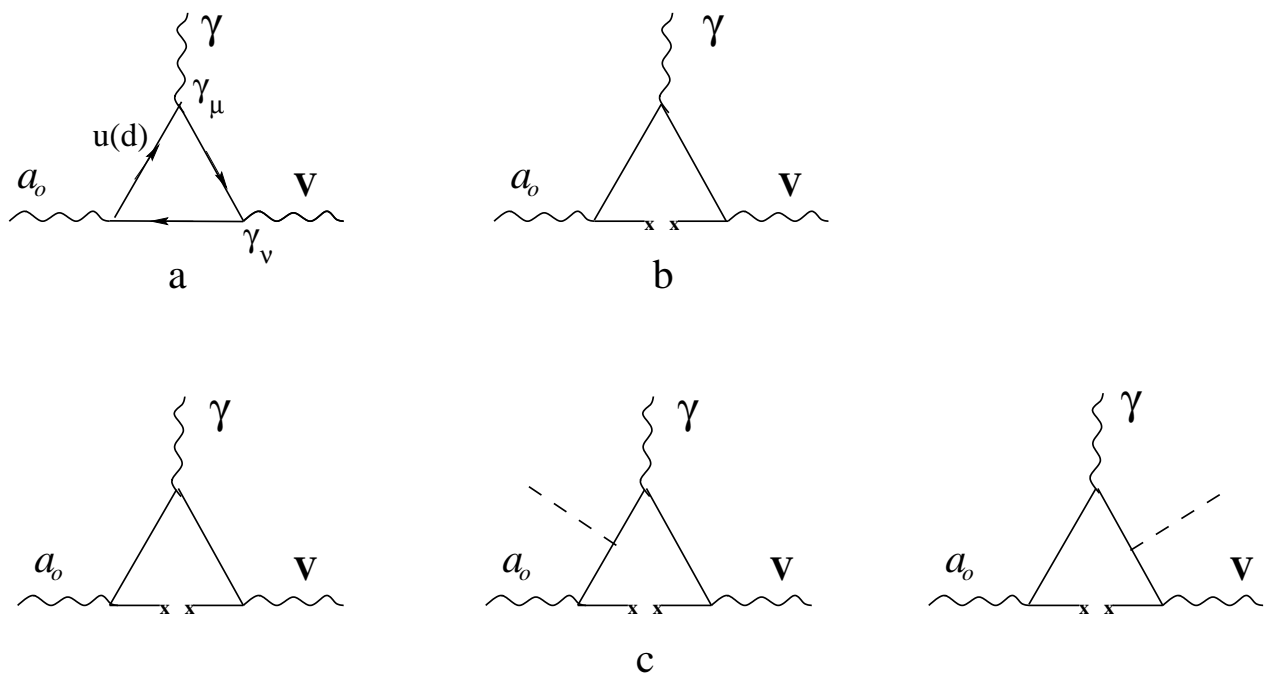
\item[{\bf Figure 2}:] Feynman Diagrams for the
$a_0V\gamma$-vertex: a- bare loop diagram, b- d=3 operator
corrections, c- d=5 operator corrections. The dotted lines denote
gluons.

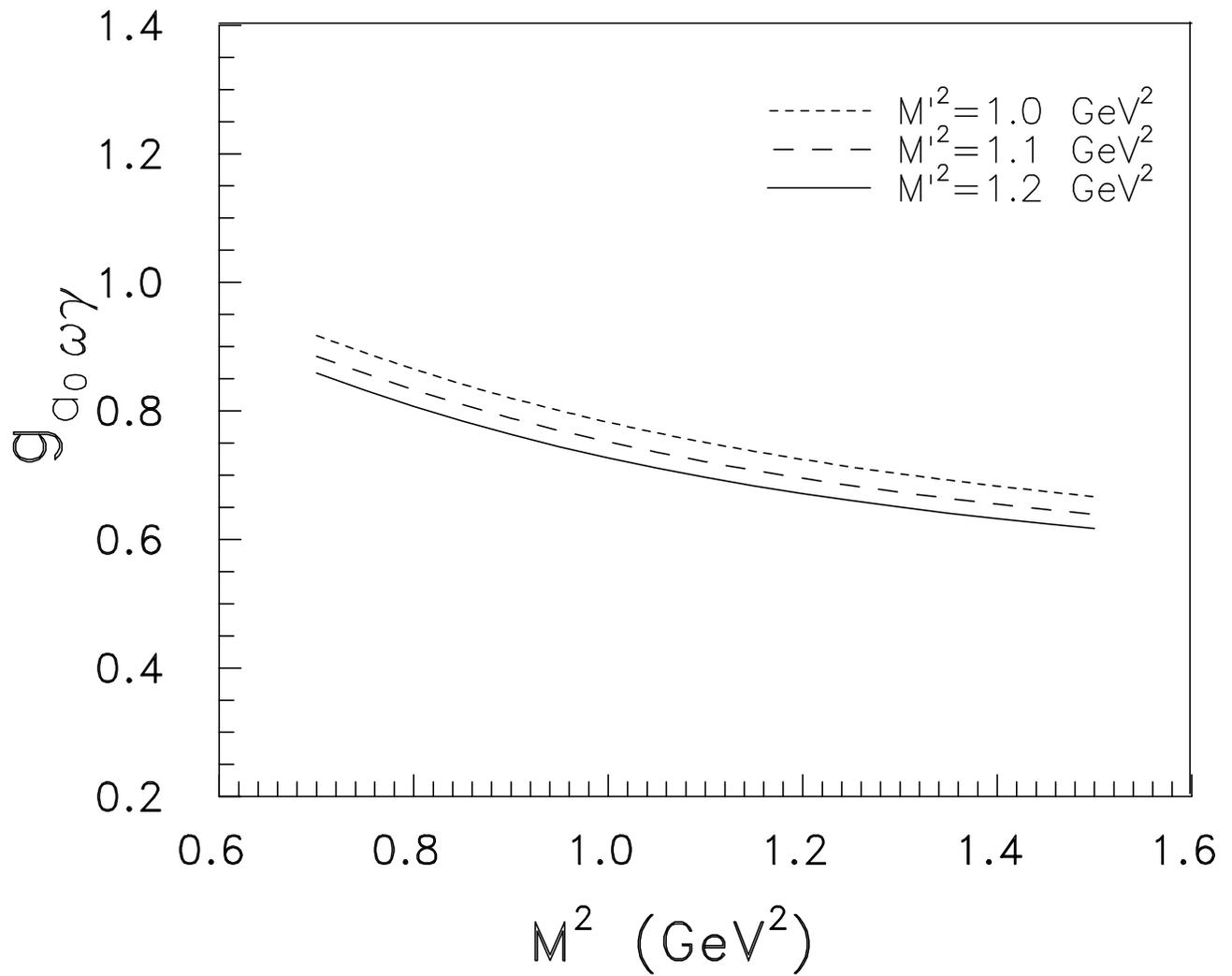
\item[{\bf Figure 3}:] The coupling constant $g_{a_0\omega\gamma}$
as a function of the Borel parameter $M^2$ for different values of
${M^\prime}^2$.

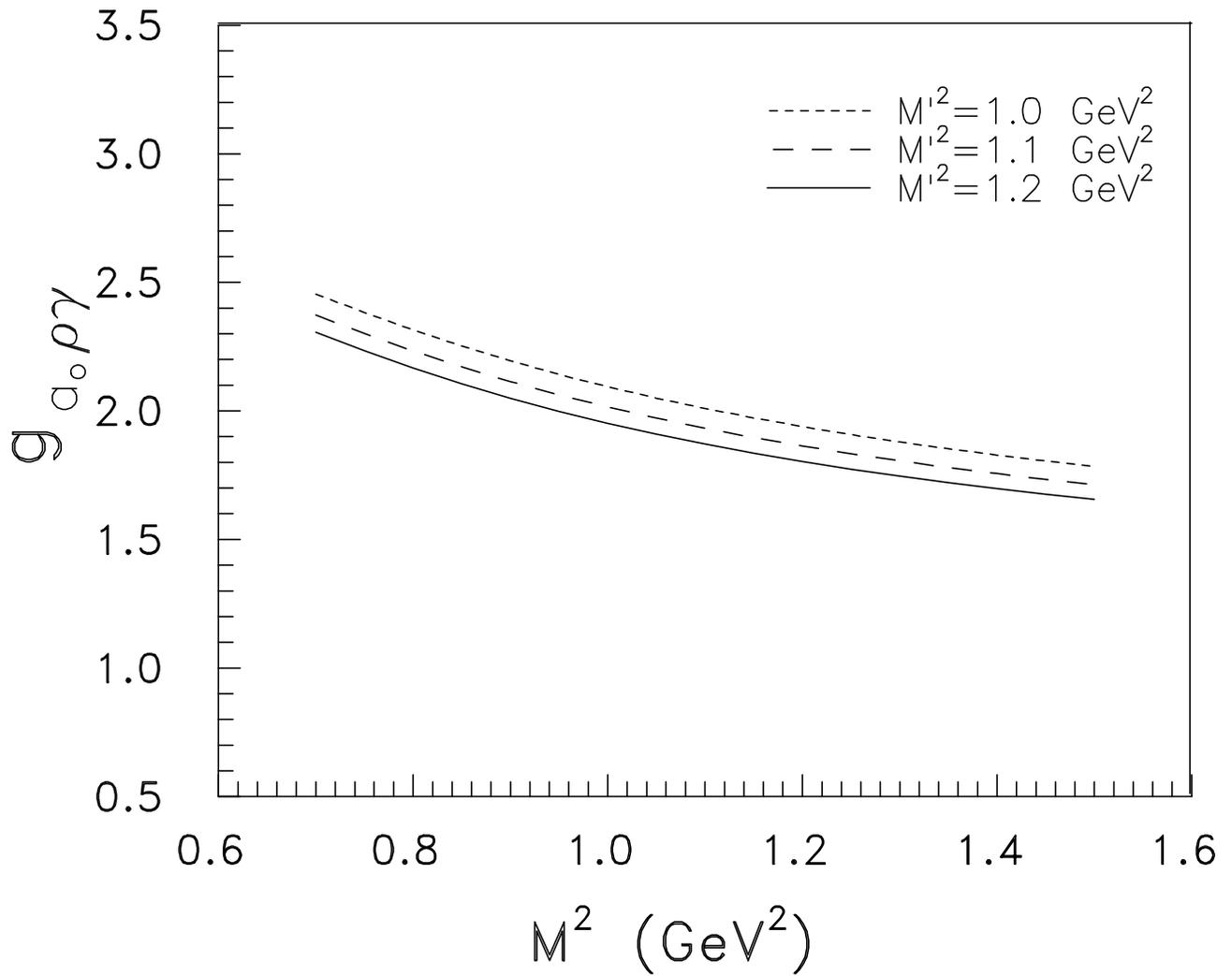
\item[{\bf Figure 4}:] The coupling constant $g_{a_0\rho\gamma}$
as a function of the Borel parameter $M^2$ for different values of
${M^\prime}^2$.
\end{description}

\end{document}